\documentclass[12pt,preprint]{aastex}

\makeatletter
  
  \@addtoreset{equation}{section}
\makeatother

\slugcomment{The Astrophysical Journal, to be submitted}

\shorttitle{The Second Born Corrections to Conductivities}
\shortauthors{Itoh et al.}

\begin{document}

\title{THE SECOND BORN CORRECTIONS TO THE ELECTRICAL AND THERMAL CONDUCTIVITIES OF DENSE MATTER IN THE LIQUID METAL PHASE}

\author{Naoki Itoh, Shinsuke Uchida, Yu Sakamoto, and Yasuharu Kohyama}
\affil{Department of Physics, Sophia University,
       7-1 Kioi-cho, Chiyoda-ku, Tokyo, 102-8554, Japan;\\
       n\_itoh@sophia.ac.jp}

\and

\author{Satoshi Nozawa}
\affil{Josai Junior College, 1-1 Keyakidai, Sakado-shi,
       Saitama, 350-0295, Japan; snozawa@josai.ac.jp}

\begin{abstract}
The second Born corrections to the electrical and thermal conductivities are calculated for the dense matter in the liquid metal phase for various elemental compositions of astrophysical importance.  Inclusion up to the second Born corrections is sufficiently accurate for the Coulomb scattering of the electrons by the atomic nuclei with $Z \lesssim 26$.  Our approach is semi-analytical, and is in contrast to that of the previous authors who have used fully numerical values of the cross section for the Coulomb scattering of the electron by the atomic nucleus.  The merit of the present semi-analytical approach is that this approach affords us to obtain the results with reliable $Z$-dependence and $\rho$-dependence.  The previous fully numerical approach has made use of the numerical values of the cross section for the scattering of the electron off the atomic nucleus for a limited number of $Z$-values, $Z$=6, 13, 29, 50, 82, and 92, and for a limited number of electron energies, 0.05MeV, 0.1MeV, 0.2MeV, 0.4MeV, 0.7MeV, 1MeV, 2MeV, 4MeV, and 10MeV.  Our study, however, has confirmed that the previous results are sufficiently accurate.  They are recovered, if the terms higher than the second Born terms are taken into account.  We make a detailed comparison of the present results with those of the previous authors.  The numerical results are parameterized in a form of analytic formulae that would facilitate practical uses of the results.  We also extend our calculations to the case of mixtures of nuclear species.  The corresponding subroutine can be retrieved from http://www.ph.sophia.ac.jp/$\tilde{ \, \, }$itoh-ken/subroutine/subroutine.htm.
\end{abstract}


\section{INTRODUCTION}

One of the present authors (N.I.) together with his collaborators has published series of papers on the calculations of the electrical and thermal conductivities of dense matter (Flowers \& Itoh 1976, 1979, 1981; Itoh et al. 1983; Mitake, Ichimaru, \& Itoh 1984; Itoh et al. 1984; Itoh \& Kohyama 1993; Itoh, Hayashi, \& Kohyama 1993).  Among these works, the calculation corresponding to the liquid metal case (Itoh et al. 1983; hereafter referred to as paper I) appears to have been most widely used in various fields of stellar evolution studies.  Therefore, it is important to keep scrutinizing the accuracy of paper I, as this paper is in frequent use among the stellar evolution researchers.

In paper I the Coulomb scatterings of the electrons off the atomic nuclei have been calculated in the framework of the first Born approximation.  Subsequently Yakovlev (1987) has made an improvement on paper I by taking into account the contributions beyond the first Born approximation.  Here we note that Yakovlev (1987) also used the analytic approach by taking into account the second Born term for the Coulomb scattering cross section.  However, his second Born corrections did not include the screening effects.  We shall consistently take into account the screening effects in our second Born corrections.  Later works by his group (Potekhin, Chabrier, \& Yakovlev 1997; Potekhin et al. 1999) improved on Yakovlev's (1987) original method by treating the first Born term and the non-Born term on the same footing, thereby taking into account the screening effects self-consistently.  In these works they have made use of the fully numerical values of the cross section for the Coulomb scattering of the electron by the atomic nucleus calculated by Doggett \& Spencer (1956).

Here we remark that the numerical calculation of the Coulomb scattering cross section by Doggett \& Spencer (1956) has been carried out for a limited number of $Z$-values for atomic nucleus $Z$=6, 13, 29, 50, 82, and 92, and for a limited number of electron energies, 0.05MeV, 0.1MeV, 0.2MeV, 0.4MeV, 0.7MeV, 1MeV, 2MeV, 4MeV, and 10MeV.  In this paper, we shall take a complementary semi-analytic approach by using the analytic expression for the second Born cross section for the Coulomb scattering of the electron off the atomic nucleus (McKinley \& Feshbach 1948; Feshbach 1952).  For nuclei $Z \lesssim 26$, the inclusion up to the second Born approximation is sufficiently accurate (Eby \& Morgan 1972).  In the following sections, however, we will confirm that the interpolations with respect to $Z$ and the electron energies done by the previous authors are remarkably accurate.

The basic formulae for the calculation of the electrical and thermal conductivities are presented in \S 2 by generalizing the formulation of paper I.  The numerical results and the assessment of the contributions beyond the first Born approximation are presented in \S 3.  The analytic formulae that fit the results of the numerical calculations are given in \S 4.  The case of the mixtures of nuclear species is dealt with in \S 5.  The last section is devoted to concluding remarks.  In the Appendix we evaluate the accuracy of the second Born approximation by comparing with the exact results obtained by Dogget \& Spencer (1956).

\section{METHOD OF CALCULATION}

We shall closely follow the method described in paper I and generalize it in such a way that it include the second Born term for the Coulomb scattering of the electron off the atomic nucleus (McKinley \& Feshbach 1948; Feshbach 1952).  The reader is referred to paper I for the earlier references in this field of research.

We shall consider the case that the atoms are completely pressure-ionized.  We further restrict ourselves to the density-temperature region in which electrons are strongly degenerate.  This condition is expressed as
\begin{eqnarray}
T & \ll & T_{F} \, = \, 5.930 \times 10^{9} \left[ \left[1 + 1.018 (Z/A)^{2/3} \rho_{6}^{2/3} \right]^{1/2} \, - \, 1 \right] \, [\rm K] \, ,
\end{eqnarray}
where $T_{F}$ is the Fermi temperature, $Z$ the atomic number of the nucleus, $A$ the mass number of the nucleus, and $\rho_{6}$ the mass density in units of 10$^{6}$ g cm$^{-3}$.  The reader is referred to the paper by Cassisi et al. (2007) for the case of the partial electron degeneracy.  For the ionic system we consider the case that it is in the liquid state.  The latest criterion corresponding to this condition is given by (Potekhin \& Chabrier 2000)
\begin{eqnarray}
1 \, \lesssim \, \Gamma & \equiv & \frac{Z^{2} e^{2}}{a k_{B} T} = 2.275 \times 10^{-1} \frac{Z^{2}}{T_{8}} \left( \frac{ \rho_{6}}{A} \right)^{1/3} \leq 175 \, ,
\end{eqnarray}
where $a=[3/(4\pi n_{i})]^{1/3}$ is the ion-sphere radius, and $T_{8}$ the temperature in units of 10$^{8}$ K.

For the calculation of the electrical and thermal conductivities we use the Ziman formula (1961) as is extended to the relativistically degenerate electrons (Flowers \& Itoh 1976).  On deriving the formula we retain the dielectric screening function due to the degenerate electrons.  As to the explicit expressions for the dielectric function, we use the relativistic formula worked out by Jancovici (1962):
\begin{eqnarray}
\epsilon(k, 0) & = & 1 + \frac{k_{TF}^{2}}{k^{2}} \left\{ \frac{2}{3} (1+b^{2})^{1/2} - \frac{2 q^{2}b}{3} {\rm sinh}^{-1}b + (1+b^{2})^{1/2} \frac{b^{2}+1-3q^{2}b^{2}}{6qb^{2}} \, {\rm ln}  \left| \frac{1+q}{1-q} \right|  \right. \nonumber \\
&  & + \left. \frac{2q^{2}b^{2}-1}{6qb^{2}} (1+q^{2}b^{2})^{1/2} \, {\rm ln} \left| \frac{q(1+b^{2})^{1/2}+(1+q^{2}b^{2})^{1/2}}{q(1+b^{2})^{1/2}-(1+q^{2}b^{2})^{1/2}} \right| \right\} \, ,
\end{eqnarray}
where $k_{TF}=(12\pi m_{e}n_{e})^{1/2}e/(\hbar k_{F})$ is the Thomas-Fermi wavenumber for the nonrelativistic electrons, $q=k/(2k_{F})$ is the momentum transfer measured in units of 2$k_{F}$, $b$ is the dimensionless relativistic parameter
\begin{eqnarray}
b & = & \frac{ \hbar k_{F}}{m_{e}c} = \frac{1}{137.036} \left( \frac{9 \pi}{4} \right)^{1/3} r_{s}^{-1} \equiv \alpha \left(\frac{9 \pi}{4} \right)^{1/3} r_{s}^{-1} \, ,
\end{eqnarray}
and $r_{s}$ is the usual electron density parameter given by
\begin{eqnarray}
r_{s} & = & 1.388 \times 10^{-2} \left( \frac{A}{Z} \right)^{1/3} \rho_{6}^{-1/3} \, .
\end{eqnarray}

Working on the transport theory for the relativistic electrons given by Flowers \& Itoh (1976) and taking into account the finite-nuclear-size corrections (Itoh \& Kohyama 1983) and the second Born term (McKinley \& Feshbach 1948; Feshbach 1952), we obtain the expression for the electrical conductivity $\sigma$ and the thermal conductivity $\kappa$:
\begin{eqnarray}
\sigma & = & 8.693 \times 10^{21} \frac{ \rho_{6}}{A} \frac{1}{ (1+b^{2})<S> } \, \left[{\rm s}^{-1} \right] \, , \\
\kappa & = & 2.363 \times 10^{17} \frac{ \rho_{6}T_{8}}{A} \frac{1}{ (1+b^{2})<S> } \, \left[ {\rm ergs \, \, cm^{-1} \, s^{-1} \, K^{-1}} \right] \, ,
\end{eqnarray}
\begin{eqnarray}
<S> & = & <S>^{1B} + <S>^{2B} \, , \\
<S>^{1B} & = & \int_{0}^{1} d \left(\frac{k}{2k_{F}} \right)  \left(\frac{k}{2k_{F}} \right)^{3} \frac{S(k/2k_{F}) \left|f(k/2k_{F}) \right|^{2}}{ \left[(k/2k_{F})^{2} \epsilon(k/2k_{F},0) \right]^{2}} \nonumber  \\
& - & \frac{b^{2}}{1 + b^{2}} \int_{0}^{1} d \left(\frac{k}{2k_{F}} \right)  \left(\frac{k}{2k_{F}} \right)^{5} \frac{S(k/2k_{F}) \left|f(k/2k_{F}) \right|^{2}}{ \left[(k/2k_{F})^{2} \epsilon(k/2k_{F},0) \right]^{2}} \nonumber \\
& \equiv & <S_{-1}> \, - \, \frac{b^{2}}{1 + b^{2}} <S_{+1}> \, , \\
<S>^{2B} & = & \pi Z \alpha \frac{b}{(1 + b^{2})^{1/2}} \nonumber  \\
& \times & \left\{ \int_{0}^{1} d \left(\frac{k}{2k_{F}} \right)  \left(\frac{k}{2k_{F}} \right)^{4} \frac{S(k/2k_{F}) \left|f(k/2k_{F}) \right|^{2}}{ \left[(k/2k_{F})^{2} \epsilon(k/2k_{F},0) \right]^{2}} \right. \nonumber  \\
&  & - \left. \int_{0}^{1} d \left(\frac{k}{2k_{F}} \right)  \left(\frac{k}{2k_{F}} \right)^{5} \frac{S(k/2k_{F}) \left|f(k/2k_{F}) \right|^{2}}{ \left[(k/2k_{F})^{2} \epsilon(k/2k_{F},0) \right]^{2}} \right\} \,  \nonumber \\
& \equiv & \pi Z \alpha \frac{b}{(1 + b^{2})^{1/2}} \left[ <S_{0}> - <S_{+1}> \right] \, ,
\end{eqnarray}
where $<S>^{1B}$ corresponds to the first Born term and $<S>^{2B}$ corresponds to the second Born term.  In the above, $\hbar k$ is the momentum transferred from the ionic system to an electron, $S(k/2k_{F})$ the ionic liquid structure factor, and $\epsilon(k/2k_{F},0)$ the static dielectric screening function due to degenerate electrons.  For the ionic liquid structure factor we use the results of Young, Corey, \& DeWitt (1991) calculated for the classical one-component plasma (OCP).  In the above formulae we have also taken into account the finite-nuclear-size corrections through the use of the atomic form factor (Itoh \& Kohyama 1983)
\begin{eqnarray}
f(q) & = & -3 \frac{(2k_{F}r_{c}q) {\rm cos}(2k_{F}r_{c}q) - {\rm sin} (2k_{F}r_{c}q)}{(2k_{F}r_{c}q)^{3}} \, ,
\end{eqnarray}
$k_{F}$ and $r_{c}$ being the electron Fermi wave number and the charge radius of the nucleus, respectively.  The electron Fermi wave number is expressed as
\begin{eqnarray}
k_{F} & = & 2.613 \times 10^{10} \left( \frac{Z}{A} \rho_{6} \right)^{1/3} \, {\rm cm}^{-1}  \, .
\end{eqnarray}
The charge radius of the nucleus is represented by
\begin{eqnarray}
r_{c} & = &  1.15 \times 10^{-13} A^{1/3} \, \, { \rm cm.}
\end{eqnarray}

The present method differs from that of Baiko et al. (1998).  These authors subtracted the contribution corresponding to the elastic scattering in the crystalline lattice phase from the total static structure factor in the liquid.  The main motivation for the modification of the structure factor near the melting point by Baiko et al. (1998) is the partial ordering of the Coulomb liquid revealed by microscopic numerical simulations.  This procedure was followed by Potekhin et al. (1999), Gnedin et al. (2001), and Cassisi et al. (2007).  In the field of condensed matter physics, however, the correctness of the original Ziman (1961) method with the use of the full liquid structure factor has long been established (Ashcroft \& Lekner 1966; Rosenfeld \& Stott 1990).

Part of the motivation for the introduction of Baiko et al.'s (1998) suggestion appears to be the finding by Itoh, Hayashi, \& Kohyama (1993) that the conductivity of astrophysical dense matter increases typically by 2--4 times upon crystallization.  Regarding this finding, we should note that simple metals in the laboratory show similar phenomena.  The electrical conductivity of the simple metals in the laboratory shows significant (2-4 times) jumps upon crystallization (Iida \& Guthrie 1993).  For these reasons we shall follow the method of paper I in which we use the full liquid structure factor, which is in accord with the method used in condensed matter physics (Ashcroft \& Lekner 1966; Rosenfeld \& Stott 1990).  Of course the analogy with simple metals should be examined with future full $ab$ $initio$ calculations.

\section{RESULTS}

We have carried out integrations in equations (2.9) and (2.10) numerically for the cases of $^{1}$H,  $^{4}$He, $^{12}$C, $^{14}$N, $^{16}$O,  $^{20}$Ne, $^{24}$Mg, $^{28}$Si, $^{32}$S, $^{40}$Ca, $^{56}$Fe by using the structure factor of the classical one-component plasma calculated by Young, Corey, \& DeWitt (1991) and Jancovici's (1962) relativistic dielectric function for degenerate electrons.  For the neutron star matter, the reader is referred to the paper by Gnedin, Yakovlev, \& Potekhin (2001).  We have made calculations for the parameter ranges, 0.1 $\leq \Gamma \leq 180$, $0 \leq {\rm log}_{10} \rho \leq 12.8$, which cover most of the density-temperature region of the dense matter in the liquid metal phase of astrophysical importance.  Note that for some elements such as the $^{56}$Fe matter these parameter ranges include the density-temperature region in which either the condition for the strong electron degeneracy or the condition for the complete pressure ionization does not hold.  All of the considered elements are certainly unstable against nuclear reactions or electron captures at extremely high densities ($\rho \gtrsim 10^{10}$gcm$^{-3}$).  We have chosen these wide parameter ranges in order to construct fitting formulae that have a wide applicable range.  The reader should use our fitting formulae in the density-temperature region in which the conditions in the above are valid.  Corresponding to the parameter range $0.1 \leq \Gamma \leq 0.2$, we have used the Debye-H$\ddot{ \rm u}$ckel form for the structure factor
\begin{eqnarray}
S(k) & = & \left[ 1 + \frac{3 \Gamma}{(ak)^{2}} \right]^{-1}  \, .
\end{eqnarray}
Here we remark that Young, Corey, \& DeWitt's (1991) calculation has been done for $\Gamma \geq 1$.  We have made a smooth extrapolation to the Debye-H$\ddot{ \rm u}$ckel regime $\Gamma \ll 1$.

In Figure 1 we show the results of the calculation for the case of $^{12}$C.  We find that the second Born corrections amount to about 2\% at $\Gamma=10$ and $\rho$=10$^{6}$g cm$^{-3}$ and about 5\% at $\Gamma=10$ and $\rho=10^{10}$g cm$^{-3}$.  In Figure 2 we show the results of the calculation for the case of $^{56}$Fe.  We find that the second Born corrections amount to about 8\% at $\Gamma=10$ and $\rho=10^{6}$g cm$^{-3}$ and about 17\% at $\Gamma=10$ and $\rho=10^{10}$g cm$^{-3}$.  These values are in good quantitative agreement with those of Potekhin, Chabrier, \& Yakovlev (1997).  For the case of $^{56}$Fe at $\Gamma=10$ and $\rho=10^{10}$g cm$^{-3}$, the present second Born corrections are significantly smaller than those of these authors who obtain about 22\% non-Born corrections for this case.  Significant part of this discrepancy appears to be due to the terms higher than the second Born term.

  In Table 1 we compare the present numerical results with the numerical results by Potekhin et al. (1997) for the cases of $\rho=10^{8}$g cm$^{-3}$; $\Gamma$=1, 10, 100.  We find generally good agreement between the present numerical results and the numerical results by Potekhin et al. (1997).  The present numerical results appear to underestimate the non-Born effects for large values of $Z$ ($Z \sim 26$).

\section{ANALYTIC FITTING FORMULAE}

We have carried out the numerical integrations of equations (2.9) and (2.10) for $^{1}$H,  $^{4}$He, $^{12}$C, $^{14}$N, $^{16}$O,  $^{20}$Ne, $^{24}$Mg, $^{28}$Si, $^{32}$S, $^{40}$Ca, $^{56}$Fe.  For the convenience of application we have fitted the numerical results of the calculation by analytic formulae.  We introduce the following variable
\begin{eqnarray}
u & = & 2 \pi ({\rm log}_{10} \rho)/25.6 \, .
\end{eqnarray}
The fitting has been carried out for the ranges $10^{0.0} \leq \rho \leq 10^{12.8}$g cm$^{-3}$, $0.1 \leq \Gamma \leq 180$.

The fitting formulae are taken as follows:
\begin{eqnarray}
<S_{-1}>(u, \Gamma) & = & v <S_{-1}>(u, 0.1) \, + \, (1-v)<S_{-1}>(u, 180) \, , \\
<S_{0}>(u, \Gamma) & = & w <S_{0}>(u, 0.1) \, + \, (1-w)<S_{0}>(u, 180) \, , \\
<S_{+1}>(u, \Gamma) & = & z <S_{+1}>(u, 0.1) \, + \, (1-z)<S_{+1}>(u, 180) \, , \\
<S_{-1}>(u, 0.1) & = & \sum_{m=1}^{5} a_{m} \, {\rm sin} \, mu \, + \, \frac{12.8}{ \pi} bu \, + \, c \, , \\
<S_{-1}>(u, 180) & = & \sum_{m=1}^{5} d_{m} \, {\rm sin} \, mu \, + \, \frac{12.8}{ \pi} eu \, + \, f \, , \\
<S_{0}>(u, 0.1) & = & \sum_{m=1}^{5} g_{m} \, {\rm sin} \, mu \, + \, \frac{12.8}{ \pi} hu \, + \, i \, , \\
<S_{0}>(u, 180) & = & \sum_{m=1}^{5} j_{m} \, {\rm sin} \, mu \, + \, \frac{12.8}{ \pi} ku \, + \, l \, , \\
<S_{+1}>(u, 0.1) & = & \sum_{m=1}^{5} p_{m} \, {\rm sin} \, mu \, + \, \frac{12.8}{ \pi} qu \, + \, r \, , \\
<S_{+1}>(u, 180) & = & \sum_{m=1}^{5} s_{m} \, {\rm sin} \, mu \, + \, \frac{12.8}{ \pi} tu \, + \, y \, , \\
v & = & \sum_{m=0}^{3} \alpha_{m} x^{m} \, , \\
w & = & \sum_{m=0}^{3} \beta_{m} x^{m} \, , \\
z & = & \sum_{m=0}^{3} \gamma_{m} x^{m} \, , \\
x & = & 0.61439 \, {\rm log}_{10} \Gamma - 0.38561 \, .
\end{eqnarray}
The coefficients are given in Tables 2--5.

The accuracy of the fitting is better than 3\% for most of the cases treated in this section.

\section{MIXTURES OF NUCLEAR SPECIES}

So far we have dealt with the case in which the matter consists of one species of atomic nucleus.  In the actual application of the present calculation to the astrophysical studies, we often encounter the case in which the matter consists of more than one species of atomic nucleus.  In this section we shall extend our calculation to the case of mixtures of nuclear species.  The case of mixtures has been discussed by Potekhin et al. (1999) and also by Brown, Bildsten, \& Chang (2000) and by Cassisi et al. (2007).  Their formalism is based on the linear mixing rule.  Here we shall give expressions according to our notations.

Let us consider the case in which the mass fraction of the nuclear species $(Z_{j}, A_{j})$ is $X_{j}$.  The electrical resistivity $R_{j}$ due to the scattering by the nuclear species $(Z_{j}, A_{j})$ is given by
\begin{eqnarray}
R_{j} & = & \frac{1+b^{2}}{8.693 \times 10^{21} \rho_{6}} \, \cdot \, \frac{X_{j} \displaystyle{ \frac{Z_{j}^{2}}{A_{j}} <S>_{j}}}{ \left( \displaystyle{\sum_{i} X_{i} \frac{Z_{i}}{A_{i}}} \right)^{2}}  \, \, \, [{\rm s}] \, ,  \\
<S>_{j} & = & <S>_{j}^{1B} + <S>_{j}^{2B} \, , \\
<S>_{j}^{1B} & = & <S_{-1}>_{j} \, - \, \frac{b^{2}}{1+b^{2}} <S_{+1}>_{j} \, ,  \\
<S>_{j}^{2B} & = & \pi Z_{j} \alpha \frac{b}{(1+b^{2})^{1/2}} \left[ <S_{0}>_{j} - <S_{+1}>_{j} \right] \, .
\end{eqnarray}
Here for the mixture case the parameter $r_{s}$ in equation (2.5) is generalized as
\begin{eqnarray}
r_{s} & = & 1.388 \times 10^{-2}  \left( \sum_{i} X_{i} \frac{Z_{i}}{A_{i}} \rho_{6} \right)^{-1/3} \, .
\end{eqnarray}
The total electrical resistivity $R$ is given by
\begin{eqnarray}
R & = & \sum_{j} R_{j}  \, .
\end{eqnarray}
Therefore, the electrical conductivity $\sigma$ is given by
\begin{eqnarray}
\sigma & = & \frac{1}{R} \, = \, \frac{8.693 \times 10^{21} \rho_{6}} {1+b^{2}} \, \cdot \, \frac{ \left( \displaystyle{ \sum_{i} X_{i} \frac{Z_{i}}{A_{i}}} \right)^{2}}{ \displaystyle{ \sum_{j} X_{j} \frac{Z_{j}^{2}}{A_{j}} <S>_{j}}}  \, \, \, [{\rm s}^{-1}] \, .
\end{eqnarray}
In the same manner, the thermal conductivity $\kappa$ is given by
\begin{eqnarray}
\kappa & = & \frac{2.363 \times 10^{17} \rho_{6} T_{8}} {1+b^{2}} \, \cdot \, \frac{ \left( \displaystyle{\sum_{i} X_{i} \frac{Z_{i}}{A_{i}}} \right)^{2}}{ \displaystyle{ \sum_{j} X_{j}  \frac{Z_{j}^{2}}{A_{j}} <S>_{j}}}  \, \, \left[ {\rm ergs \, \, cm^{-1} \, s^{-1} \, K^{-1}} \right] \, .
\end{eqnarray}
In the above, the scattering integral $<S>_{j}$ corresponding to the nuclear species $(Z_{j}, A_{j})$ should be calculated by using the Coulomb coupling parameter (Itoh et al. 1979; Potekhin et al. 1999; Brown, Bildsten, \& Chang 2002; Itoh et al. 2004)
\begin{eqnarray}
\Gamma_{j} & = & \frac{Z_{j}^{5/3} e^{2}}{a_{e} k_{B}T} \, = \, 0.2275 \frac{Z_{j}^{5/3}}{T_{8}} \left( \sum_{i} X_{i} \frac{Z_{i}}{A_{i}} \rho_{6} \right)^{1/3} \, ,  \\
a_{e} & = & \left( \frac{3}{4 \pi n_{e}} \right)^{1/3} = \left( \frac{3}{4 \pi \displaystyle{ \sum_{i} n_{i} Z_{i}}} \right)^{1/3}  \, ,
\end{eqnarray}
where $a_{e}$ is the electron-sphere radius, and $n_{e}$ and $n_{i}$ are the number densities of the electrons and the $i$-th nuclear species $(Z_{i}, A_{i})$, respectively.

\section{CONCLUDING REMARKS}

We have calculated the second Born corrections to the electrical and thermal conductivities of the dense matter in the liquid metal phase for various elemental compositions of astrophysical importance by extending the calculations reported in paper I.  We have used the semi-analytical approach which is in contrast to that of the previous authors (Yakovlev 1987; Potekhin, Chabrier, \& Yakovlev 1997; Potekhin et al. 1999), who made use of the fully numerical values of the cross section for the scattering of the electron by the atomic nucleus calculated by Doggett \& Spencer (1956).  It should be noted that the numerical calculation of the Coulomb scattering cross section by Doggett \& Spencer (1956) has been carried out for a limited number of $Z$-values for the atomic nucleus $Z$=6, 13, 29, 50, 82, and 92, and for a limited number of electron energies 0.05MeV, 0.1MeV, 0.2MeV, 0.4MeV, 0.7MeV, 1MeV, 2MeV, 4MeV, and 10MeV, and also for a limited number (13) of the scattering angles that are related to $k/2k_{F}$ in equations (2.9) and (2.10).  The sparseness of data for light and medium nuclei (only for $Z$=6, 13, 29) is potentially vulnerable in order to obtain results with reliable $Z$-dependence.  However, our study has confirmed that the previous results have sufficiently accurate $Z$-dependence and $\rho$-dependence, since they are recovered, within about 1\%, if our second-Born results are multiplied by the ratio of the full non-Born $<S>^{DS}$ to the second-Born $<S>^{1B+2B}$.  The definitions of $<S>^{DS}$ and $<S>^{1B+2B}$ are given in the Appendix.

We have found that our results are in general agreement with those of Potekhin, Chabrier, \& Yakovlev (1997).  Our second Born corrections are significantly smaller than the non-Born corrections of these authors for the case of $^{56}$Fe at $\Gamma=10$ and $\rho=10^{10}$g cm$^{-3}$.  Significant part of this discrepancy appears to be due to the terms higher than the second Born term.

In the present calculation, in contrast to Baiko et al. (1998), we have used the full liquid structure factor, for the reasons explained in \S 2.

We have summarized our numerical results by accurate analytic fitting formulae.  We have also presented the prescriptions to deal with the cases of mixtures of nuclear species.  Therefore, the present results should be readily applied to various studies in the field of stellar evolution.

\acknowledgements
We wish to thank our referee for many useful comments that have greatly helped us in revising the manuscript.  We also wish to thank D. G. Yakovlev and A. Y. Potekhin for their very informative communication and providing us with the numerical data of their results in Table 1.  One of the authors (N.I.) wishes to thank N. W. Ashcroft, K. Hoshino, and H. Maebashi for their expert advice regarding the calculations of the conductivities of simple metals in the laboratory.  He especially appreciates N. W. Ashcroft's lucid reasoning regarding the correctness of Ziman's original method with the use of the full liquid structure factor.  He also wishes to thank H. E. DeWitt and S. Hansen for their valuable communication regarding the OCP structure factor.  This work is financially supported in part by the Grant-in-Aid for Scientific Research of Japanese Ministry of Education, Culture, Sports, Science, and Technology under the contract 16540220.

\appendix
\section{APPENDIX}

  In this Appendix we evaluate the accuracy of the second Born approximation by comparing with the exact results obtained by Doggett \& Spencer (1956).  The second Born approximation gives a correction factor to the Rutherford cross section (McKinley \& Feshbach 1948; Feshbach 1952):
\begin{eqnarray}
R^{1B+2B} & = & 1 - \beta^{2} \, {\rm sin}^{2} \frac{\theta}{2} + \pi Z \alpha \beta \, {\rm sin} \frac{\theta}{2} \left(1 - {\rm sin} \frac{\theta}{2} \right)  \, ,
\end{eqnarray}
where
\begin{eqnarray}
\beta & = & \frac{\left[ (E_{kin}/0.5110{\rm MeV})^{2} + 2 (E_{kin}/0.5110{\rm MeV}) \right]^{1/2}}{1 + (E_{kin}/0.5110{\rm MeV})}  \, ,
\end{eqnarray}
$E_{kin}$ being the kinetic energy of the electron, and $\theta$ is the angle of scattering.  The $<S>$ factor corresponding to the results by Doggett \& Spencer (1956) is defined by
\begin{eqnarray}
<S>^{DS} & \equiv & \int_{0}^{1} d \left(\frac{k}{2k_{F}} \right)  \left(\frac{k}{2k_{F}} \right)^{3} \frac{S(k/2k_{F}) \left|f(k/2k_{F}) \right|^{2}}{ \left[(k/2k_{F})^{2} \epsilon(k/2k_{F},0) \right]^{2}} \, R^{DS}(E_{kin}, k/2k_{F})  \, ,
\end{eqnarray}
where $k$ is related to $\theta$ by
\begin{eqnarray}
\frac{k}{2 k_{F}} & = & {\rm sin} \frac{\theta}{2}  \, .
\end{eqnarray}
In order to make the comparison self-consistent, in this Appendix we define
\begin{eqnarray}
<S>^{1B+2B} & \equiv & \int_{0}^{1} d \left(\frac{k}{2k_{F}} \right)  \left(\frac{k}{2k_{F}} \right)^{3} \frac{S(k/2k_{F}) \left|f(k/2k_{F}) \right|^{2}}{ \left[(k/2k_{F})^{2} \epsilon(k/2k_{F},0) \right]^{2}} \, R^{1B+2B}  \nonumber  \\
& = & <S>^{1B} + <S>^{2B}  \, ,
\end{eqnarray}
which of course coincides with our previous equations (2.8), (2.9), (2.10).  Here we have used the relationship
\begin{eqnarray}
E_{kin} & = & 0.5110{\rm MeV} \left\{ \left[1 + 1.018 (Z/A)^{2/3} \rho_{6}^{2/3} \right]^{1/2} \, - \, 1 \right\} \, .
\end{eqnarray}

  In Table 6 we compare the results corresponding to the second Born approximation with those corresponding to Doggett \& Spencer (1956) for the cases of $\Gamma$=10; $Z$=6, 13, 29; and $E_{kin}$=0.05MeV, 0.1MeV, 0.2MeV, 0.4MeV, 0.7MeV, 1MeV, 2MeV, 4MeV, 10MeV.  We find the accuracy of the second Born correction is better than 0.4\% for $Z$=6, better than 1.4\% for $Z$=13, and better than 6.0\% for $Z$=29.

\renewcommand{\baselinestretch}{1.0}

\clearpage

\begin{deluxetable}{cccc}
\tablecaption{Comparison of the present numerical results with the numerical results by Potekhin et al. (1997) for the cases of $\rho=10^{8}$g cm$^{-3}$; $\Gamma$=1, 10, 100.}
\tablewidth{0pt}
\tablehead{
\colhead{$\Gamma$} & \colhead{$Z$}    & \colhead{$<S>^{ \rm present}$}  & 
\colhead{$<S>^{ \rm Potekhin \, et \, al.}$}
}
\startdata
    1  &   6  &  1.0841  &  1.087\\
       &   7  &  1.1297  &  1.133\\
       &   8  &  1.1701  &  1.175\\
       &  10  &  1.2400  &  1.248\\
       &  12  &  1.2996  &  1.311\\
       &  14  &  1.3521  &  1.368\\
       &  16  &  1.3995  &  1.420\\
       &  20  &  1.4834  &  1.516\\
       &  26  &  1.5925  &  1.649\\ \hline
  10   &   6  &  0.6490  &  0.651\\
       &   7  &  0.6975  &  0.701\\
       &   8  &  0.7407  &  0.745\\
       &  10  &  0.8159  &  0.823\\
       &  12  &  0.8806  &  0.891\\
       &  14  &  0.9379  &  0.953\\
       &  16  &  0.9898  &  1.009\\
       &  20  &  1.0819  &  1.113\\
       &  26  &  1.2017  &  1.256\\ \hline
 100   &   6  &  0.5236  &  0.526\\
       &   7  &  0.5717  &  0.575\\
       &   8  &  0.6152  &  0.620\\
       &  10  &  0.6929  &  0.700\\
       &  12  &  0.7609  &  0.771\\
       &  14  &  0.8209  &  0.836\\
       &  16  &  0.8750  &  0.880\\
       &  20  &  0.9700  &  1.001\\
       &  26  &  1.0931  &  1.148\\
\enddata
\end{deluxetable}
\clearpage

\begin{deluxetable}{crrrrrrrrrrr}
\tabletypesize{\scriptsize}
\rotate
\tablecaption{Coefficients in the fitting formulae for $<S_{-1}>(u, 0.1)$ and $<S_{-1}>(u, 180)$}
\tablewidth{0pt}
\tablehead{
\colhead{Coefficient} & \colhead{$^{1}$H}    & \colhead{$^{4}$He}  & 
\colhead{$^{12}$C}    & \colhead{$^{14}$N}   & \colhead{$^{16}$O}  & 
\colhead{$^{20}$Ne}   & \colhead{$^{24}$Mg}  & \colhead{$^{28}$Si} & 
\colhead{$^{32}$S}    & \colhead{$^{40}$Ca}  & \colhead{$^{56}$Fe} 
}
\startdata
$a_{1}$ & 0.6496 & 0.7407 & 0.8981 & 0.9232 & 0.9457 & 0.9848 & 1.0181 & 1.0471 &	1.0729 & 1.1171	& 1.1690\\
$a_{2}$ & 0.0471 & $-$0.0007 & $-$0.0666 & $-$0.0781 & $-$0.0884 & $-$0.1065	& $-$0.1221	& $-$0.1357	& $-$0.1477	& $-$0.1684	& $-$0.1970\\
$a_{3}$ & $-$0.0056 & $-$0.0165 & $-$0.0071 & $-$0.0045 & $-$0.0019 & 0.0031	& 0.0076 & 0.0117 &	0.0155 & 0.0222	& 0.0297\\
$a_{4}$ & $-$0.0284 & $-$0.0376 & $-$0.0558 & $-$0.0588 & $-$0.0615 & $-$0.0663	& $-$0.0703	& $-$0.0737	& $-$0.0767	& $-$0.0818	& $-$0.0869\\
$a_{5}$ & 0.0054 & 0.0114 & 0.0247 & 0.0270 & 0.0291 & 0.0326 & 0.0356 & 0.0382 &	0.0404 & 0.0440 & 0.0481\\
$b$     & 0.0921 & 0.1037 & 0.1068 & 0.1064 & 0.1059 & 0.1046 & 0.1032 & 0.1018 &	0.1004 & 0.0977	& 0.0946\\
$c$     & 0.4531 & 0.3959 & 0.4040 & 0.4047 & 0.4053 & 0.4063 & 0.4069 & 0.4074 &	0.4078 & 0.4084	& 0.4017\\
$d_{1}$	& 0.0268 & 0.2196 & 0.4347 & 0.4753 & 0.5166 & 0.5930 &	0.6520 & 0.6976 &	0.7358 & 0.8004 & 0.8856\\
$d_{2}$	& 0.0012 & 0.0006 & 0.0084 & 0.0052 & 0.0006 & $-$0.0094 & $-$0.0176	&	$-$0.0243 & $-$0.0304 &	$-$0.0423 & $-$0.0612\\
$d_{3}$	& 0.0051 & 0.0440 & 0.0741 & 0.0796 & 0.0854 & 0.0962 &	0.1035 & 0.1082 &	0.1116 & 0.1169 & 0.1224\\
$d_{4}$	& $-$0.0007 & $-$0.0081	& $-$0.0189 & $-$0.0224	& $-$0.0263 & $-$0.0338	&	$-$0.0399 & $-$0.0449 &	$-$0.0494 & $-$0.0571 &	$-$0.0679\\
$d_{5}$	& 0.0018 & 0.0155 & 0.0228 & 0.0241 & 0.0257 & 0.0287 &	0.0305 & 0.0316 &	0.0323 & 0.0333 & 0.0344\\
$e$	& 0.0011 & 0.0056 & 0.0174 & 0.0184 & 0.0189 & 0.0194 &	0.0200 & 0.0208 &	0.0214 & 0.0223 & 0.0230\\
$f$	& 0.0621 & 0.3641 & 0.3604 & 0.3670 & 0.3787 & 0.4011 &	0.4097 & 0.4097 &	0.4074 & 0.4036 & 0.3986\\
\enddata
\end{deluxetable}

\clearpage

\begin{deluxetable}{crrrrrrrrrrr}
\tabletypesize{\scriptsize}
\rotate
\tablecaption{Coefficients in the fitting formulae for $<S_{0}>(u, 0.1)$ and $<S_{0}>(u, 180)$}
\tablewidth{0pt}
\tablehead{
\colhead{Coefficient} & \colhead{$^{1}$H}    & \colhead{$^{4}$He}  & 
\colhead{$^{12}$C}    & \colhead{$^{14}$N}   & \colhead{$^{16}$O}  & 
\colhead{$^{20}$Ne}   & \colhead{$^{24}$Mg}  & \colhead{$^{28}$Si} & 
\colhead{$^{32}$S}    & \colhead{$^{40}$Ca}  & \colhead{$^{56}$Fe} 
}
\startdata
$g_{1}$	& 0.2781 & 0.3281 &	0.4042 & 0.4170 & 0.4286 & 0.4489 & 0.4662 & 0.4813 &	0.4946 & 0.5173	& 0.5452\\
$g_{2}$	& 0.0357 & 0.0222 &	$-$0.0077 &	$-$0.0131 &	$-$0.0180 & $-$0.0266 & $-$0.0339	& $-$0.0404	& $-$0.0460 & $-$0.0556 & $-$0.0672\\
$g_{3}$	& 0.0224 & 0.0249 &	0.0396 & 0.0423	& 0.0448 & 0.0491 & 0.0528 & 0.0560 & 0.0588 & 0.0635 & 0.0684\\
$g_{4}$	& $-$0.0072	& $-$0.0134	& $-$0.0245	& $-$0.0264	& $-$0.0280 & $-$0.0309 & $-$0.0333 & $-$0.0353	& $-$0.0370 & $-$0.0397 & $-$0.0426\\
$g_{5}$	& 0.0059 & 0.0082 &	0.0152 & 0.0163	& 0.0174 & 0.0191 & 0.0205 & 0.0217 &	0.0227 & 0.0241	& 0.0254\\
$h$	& 0.0303 & 0.0323 &	0.0269 & 0.0258	& 0.0247 & 0.0227 & 0.0209 & 0.0193 &	0.0179 & 0.0154	& 0.0127\\
$i$	& 0.3087 & 0.2749 &	0.2790 & 0.2794 & 0.2797 & 0.2802 & 0.2805 & 0.2808 &	0.2810 & 0.2813 & 0.2773\\
$j_{1}$	& 0.0225 & 0.1881 & 0.2913 & 0.3121 & 0.3353 & 0.3787 &	0.4085 & 0.4283 &	0.4436 & 0.4687 & 0.5053\\
$j_{2}$	& 0.0007 & $-$0.0008 & 0.0011 &	$-$0.0026 & $-$0.0072 &	$-$0.0166	&	$-$0.0233 & $-$0.0281 & $-$0.0321 & $-$0.0395 &	$-$0.0505\\
$j_{3}$	& 0.0045 & 0.0385 & 0.0519 & 0.0551 & 0.0590 & 0.0664 & 0.0711 & 0.0736 &	0.0754 & 0.0781 & 0.0819\\
$j_{4}$	& $-$0.0006 & $-$0.0071 & $-$0.0133 & $-$0.0156 & $-$0.0182 & $-$0.0229	&	$-$0.0263 & $-$0.0288 &	$-$0.0307 & $-$0.0339 &	$-$0.0383\\
$j_{5}$	& 0.0016 & 0.0136 & 0.0164 & 0.0173 & 0.0185 & 0.0208 &	0.0221 & 0.0228 &	0.0231 & 0.0236 & 0.0243\\
$k$	& 0.0007 & 0.0041 & 0.0095 & 0.0093 & 0.0087 & 0.0074 &	0.0066 & 0.0061 &	0.0056 & 0.0047 & 0.0033\\
$l$	& 0.0557 & 0.3200 & 0.2571 & 0.2604 & 0.2694 & 0.2873 &	0.2923 & 0.2897 &	0.2856 & 0.2799 & 0.2763\\
\enddata
\end{deluxetable}

\clearpage

\begin{deluxetable}{crrrrrrrrrrr}
\tabletypesize{\scriptsize}
\rotate
\tablecaption{Coefficients in the fitting formulae for $<S_{+1}>(u, 0.1)$ and $<S_{+1}>(u, 180)$}
\tablewidth{0pt}
\tablehead{
\colhead{Coefficient} & \colhead{$^{1}$H}    & \colhead{$^{4}$He}  & 
\colhead{$^{12}$C}    & \colhead{$^{14}$N}   & \colhead{$^{16}$O}  & 
\colhead{$^{20}$Ne}   & \colhead{$^{24}$Mg}  & \colhead{$^{28}$Si} & 
\colhead{$^{32}$S}    & \colhead{$^{40}$Ca}  & \colhead{$^{56}$Fe} 
}
\startdata
$p_{1}$	& 0.1543 & 0.1881 &	0.2380 & 0.2466 & 0.2544 & 0.2679 & 0.2794 & 0.2893 &	0.2980 & 0.3126 & 0.3306\\
$p_{2}$	& 0.0202 & 0.0137 &	$-$0.0068 &	$-$0.0105 &	$-$0.0138 &	$-$0.0196 & $-$0.0244	& $-$0.0287	& $-$0.0323	& $-$0.0384	& $-$0.0451\\
$p_{3}$	& 0.0205 & 0.0248 &	0.0367 & 0.0388	& 0.0406 & 0.0438 & 0.0465 & 0.0487 &	0.0506 & 0.0537	& 0.0569\\
$p_{4}$	& $-$0.0024	& $-$0.0063	& $-$0.0140 & $-$0.0152 & $-$0.0164 & $-$0.0183	& $-$0.0198	& $-$0.0211	& $-$0.0222	& $-$0.0238	& $-$0.0254\\
$p_{5}$	& 0.0057 & 0.0074 &	0.0123 & 0.0131	& 0.0137 & 0.0149 & 0.0158 & 0.0165 &	0.0170 & 0.0178 & 0.0183\\
$q$	& 0.0133 & 0.0135 &	0.0084 & 0.0074	& 0.0065 & 0.0049 & 0.0036 & 0.0024 &	0.0013 & $-$0.0005 & $-$0.0024\\
$r$	& 0.2293 & 0.2069 &	0.2095 & 0.2097	& 0.2099 & 0.2102 & 0.2104 & 0.2105 &	0.2106 & 0.2108	& 0.2082 \\
$s_{1}$	& 0.0196 & 0.1636 & 0.2027 & 0.2148 & 0.2302 & 0.2597 &	0.2776 & 0.2872 &	0.2938 & 0.3049 & 0.3248\\
$s_{2}$	& 0.0004 & $-$0.0017 & $-$0.0028 & $-$0.0064 & $-$0.0108 & $-$0.0191	&	$-$0.0245 & $-$0.0279 &	$-$0.0305 & $-$0.0352 & $-$0.0422\\
$s_{3}$	& 0.0041 & 0.0339 & 0.0375 & 0.0397 & 0.0428 & 0.0487 &	0.0521 & 0.0536 &	0.0545 & 0.0561 & 0.0589\\
$s_{4}$	& $-$0.0006 & $-$0.0064	& $-$0.0099 & $-$0.0115	& $-$0.0134 & $-$0.0168	&	$-$0.019 & $-$0.0202 & $-$0.0211 & $-$0.0226 & $-$0.0246\\
$s_{5}$	& 0.0015 & 0.0121 & 0.0121 & 0.0129 & 0.0139 & 0.0159 &	0.0170 & 0.0174 &	0.0175 & 0.0177 & 0.0181\\
$t$	& 0.0005 & 0.0032 & 0.0051 & 0.0044 & 0.0035 & 0.0017 &	0.0005 & $-$0.0002	& $-$0.0008 & $-$0.0019	& $-$0.0035\\
$y$	& 0.0507 & 0.2834 & 0.1902 & 0.1933 & 0.2021 & 0.2188 &	0.2226 & 0.2192 &	0.2147 & 0.2089 & 0.2068\\
\enddata
\end{deluxetable}

\clearpage

\begin{deluxetable}{crrrrrrrrrrr}
\tabletypesize{\scriptsize}
\rotate
\tablecaption{Coefficients in the fitting formulae for $v$, $w$, and $z$}
\tablewidth{0pt}
\tablehead{
\colhead{Coefficient} & \colhead{$^{1}$H}    & \colhead{$^{4}$He}  & 
\colhead{$^{12}$C}    & \colhead{$^{14}$N}   & \colhead{$^{16}$O}  & 
\colhead{$^{20}$Ne}   & \colhead{$^{24}$Mg}  & \colhead{$^{28}$Si} & 
\colhead{$^{32}$S}    & \colhead{$^{40}$Ca}  & \colhead{$^{56}$Fe}
}
\startdata
$\alpha_{0}$ & 0.4288 &	0.1778 & 0.2634 & 0.2661 & 0.2619 & 0.2489 & 0.2461 	& 0.2513 & 0.2583 & 0.2700 & 0.2760\\
$\alpha_{1}$ & $-$0.5654 & $-$0.5446 & $-$0.5073 & $-$0.5144 & $-$0.5251	& $-$0.5454 & $-$0.5525	& $-$0.5517 & $-$0.5495	& $-$0.5480 & $-$0.5561\\
$\alpha_{2}$ & 0.0769 &	0.3337 & 0.2504 & 0.2474 & 0.2512 & 0.2638 & 0.2674 	& 0.2632 & 0.2567 & 0.2453 & 0.2389\\
$\alpha_{3}$ & 0.0662 &	0.0467 & 0.0083 & 0.0155 & 0.0264 & 0.0472 & 0.0549 	& 0.0546 & 0.0527 & 0.0512 & 0.0594\\
$\beta_{0}$ & 0.5283 & 0.0090 &	0.3358 & 0.3352 & 0.3056 & 0.2147 & 0.1748 	& 0.1905 & 0.2221 & 0.2725 & 0.2749\\
$\beta_{1}$ & $-$0.5546	& $-$0.7314 & $-$0.4548	& $-$0.4723 & $-$0.5089 & $-$0.5916 & $-$0.6101	& $-$0.5802 & $-$0.5437	& $-$0.5001 & $-$0.5121\\
$\beta_{2}$ & $-$0.0256	& 0.5152 & 0.1868 & 0.1849 & 0.2120 & 0.3007 &	0.3446 	& 0.3334 & 0.3045 & 0.2539 & 0.2464\\
$\beta_{3}$ & 0.0558 & 0.2428 &	$-$0.0421 & $-$0.0245 &	0.0124 & 0.0973 & 0.1193& 0.0915 & 0.0555 & 0.0106 & 0.0208\\
$\gamma_{0}$ & 0.6074 &	$-$4.1967 & 0.4740 & 0.4663 & 0.3642 & $-$0.2350 & $-$0.9551 & $-$0.6418 & $-$0.1656 & 0.2635 &	0.2676\\
$\gamma_{1}$ & $-$0.5134 & $-$2.8194 & $-$0.1732 & $-$0.2319 & $-$0.3358 & $-$0.7496 & $-$1.0234 & $-$0.6404 & $-$0.3168 & $-$0.1219 & $-$0.1927\\
$\gamma_{2}$ & $-$0.1064 & 4.9319 & 0.0621 & 0.0612 & 0.1539 & 0.7369 &	1.4929 	& 1.2219 & 0.7513 & 0.2992 & 0.2587\\
$\gamma_{3}$ & 0.0142 &	2.4604 & $-$0.3263 & $-$0.2676 & $-$0.1633 & 0.2629 & 0.5802 & 0.2094 &	$-$0.1265 & $-$0.3460 &	$-$0.2894\\
\enddata
\end{deluxetable}

\clearpage

\begin{deluxetable}{ccccc}
\tablecaption{Comparison of the results corresponding to the second Born approximation with those corresponding to Doggett \& Spencer (1956) for the cases of $\Gamma$=10; $Z$=6, 13, 29; and $E_{kin}$=0.05MeV, 0.1MeV, 0.2MeV, 0.4MeV, 0.7MeV, 1MeV, 2MeV, 4MeV, 10MeV.}
\tablewidth{0pt}
\tablehead{
\colhead{$Z$}  &  \colhead{$E_{kin}$(MeV)}  &  \colhead{$<S>^{DS}$}  &  \colhead{$<S>^{1B+2B}$}  &  \colhead{$<S>^{1B+2B}/<S>^{DS}$}
}
\startdata
6	&	0.05	&	0.9324 	&	0.9299 	&	0.9973\\
	&	0.1	&	0.8921 	&	0.8897 	&	0.9973\\
	&	0.2	&	0.8263 	&	0.8239 	&	0.9971\\
	&	0.4	&	0.7489 	&	0.7465 	&	0.9969\\
	&	0.7	&	0.6950 	&	0.6929 	&	0.9970\\
	&	1	&	0.6698 	&	0.6677 	&	0.9969\\
	&	2	&	0.6410 	&	0.6386 	&	0.9963\\
	&	4	&	0.6288 	&	0.6268 	&	0.9968\\
	&	10	&	0.6224 	&	0.6202 	&	0.9965\\ \hline
13	&	0.05	&	1.1805 	&	1.1668 	&	0.9883\\
	&	0.1	&	1.1506 	&	1.1365 	&	0.9878\\
	&	0.2	&	1.0914 	&	1.0778 	&	0.9876\\
	&	0.4	&	1.0182 	&	1.0048 	&	0.9868\\
	&	0.7	&	0.9658 	&	0.9529 	&	0.9867\\
	&	1	&	0.9412 	&	0.9283 	&	0.9863\\
	&	2	&	0.9125 	&	0.8998 	&	0.9860\\
	&	4	&	0.9008 	&	0.8879 	&	0.9857\\
	&	10	&	0.8915 	&	0.8788 	&	0.9858\\ \hline
29	&	0.05	&	1.5000 	&	1.4295 	&	0.9530\\
	&	0.1	&	1.4999 	&	1.4239 	&	0.9493\\
	&	0.2	&	1.4664 	&	1.3875 	&	0.9462\\
	&	0.4	&	1.4113 	&	1.3321 	&	0.9439\\
	&	0.7	&	1.3685 	&	1.2900 	&	0.9426\\
	&	1	&	1.3485 	&	1.2695 	&	0.9414\\
	&	2	&	1.3240 	&	1.2453 	&	0.9405\\
	&	4	&	1.3125 	&	1.2343 	&	0.9405\\
	&	10	&	1.2983 	&	1.2204 	&	0.9400\\
\enddata
\end{deluxetable}
\clearpage

\begin{figure}
\epsscale{.80}
\plotone{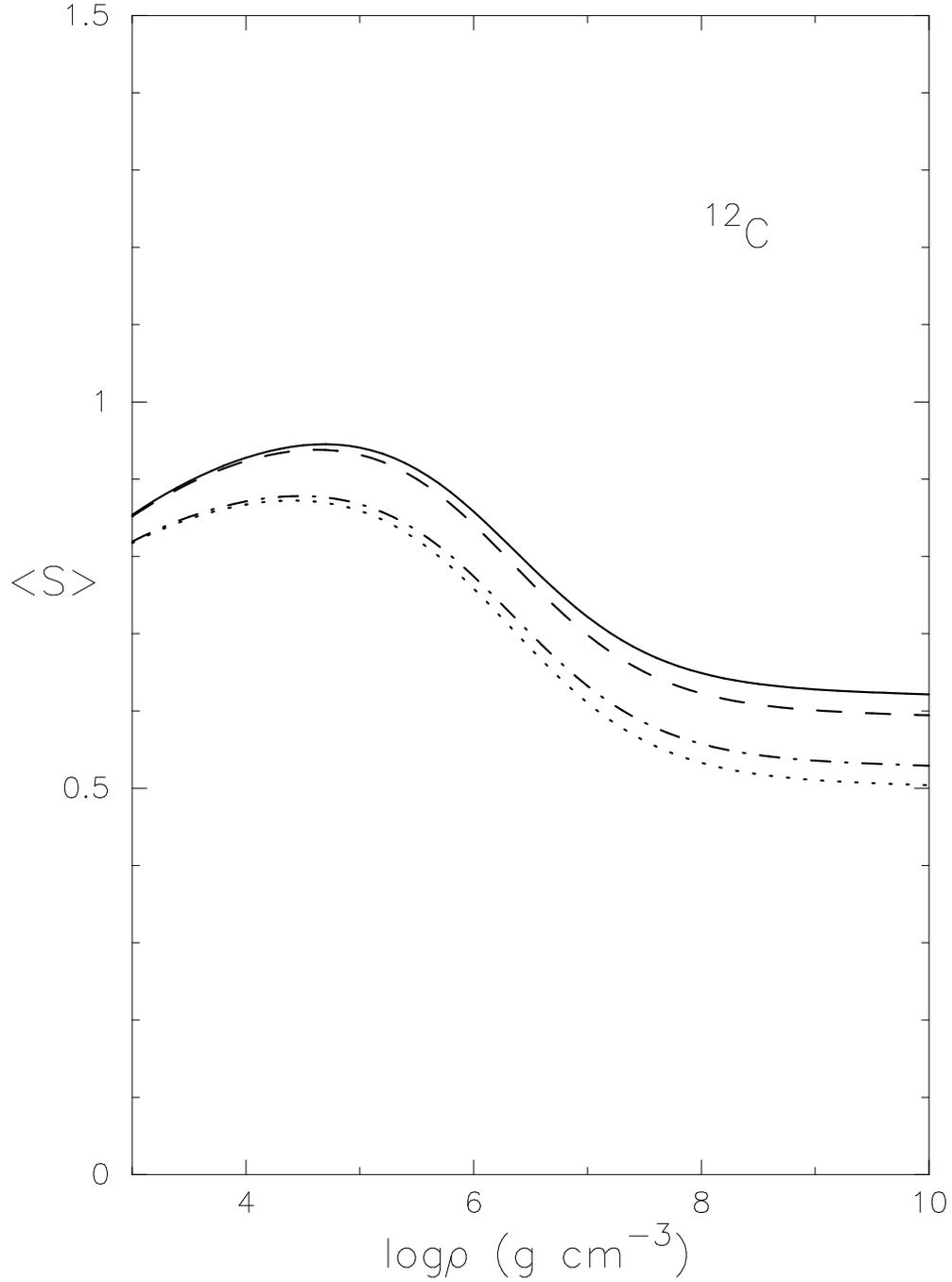}
\caption{The values of $<S>^{1B}+<S>^{2B}$ and $<S>^{1B}$ for the $^{12}$C matter.  The solid curve is the contribution of $<S>^{1B}+<S>^{2B}$ for $\Gamma=10$.  The dashed curve is the contribution of $<S>^{1B}$ for $\Gamma=10$.  The dash-dotted curve is the contribution of $<S>^{1B}+<S>^{2B}$ for $\Gamma=40$.  The dotted curve is the contribution of $<S>^{1B}$ for $\Gamma=40$.}
\end{figure}

\clearpage

\begin{figure}
\epsscale{.80}
\plotone{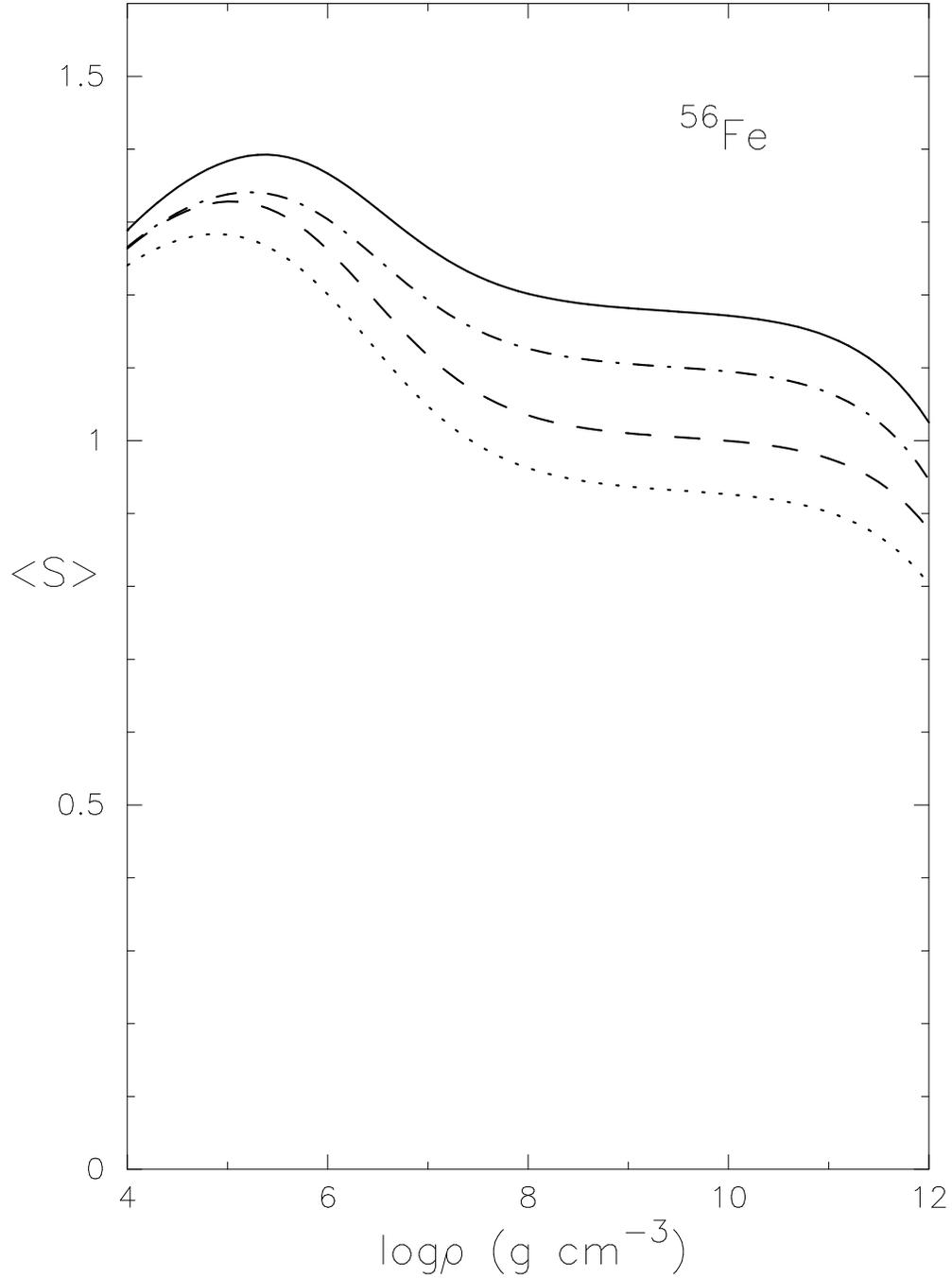}
\caption{The values of $<S>^{1B}+<S>^{2B}$ and $<S>^{1B}$ for the $^{56}$Fe matter.  The solid curve is the contribution of $<S>^{1B}+<S>^{2B}$ for $\Gamma=10$.  The dashed curve is the contribution of $<S>^{1B}$ for $\Gamma=10$.  The dash-dotted curve is the contribution of $<S>^{1B}+<S>^{2B}$ for $\Gamma=40$.  The dotted curve is the contribution of $<S>^{1B}$ for $\Gamma=40$.}
\end{figure}

\end{document}